%% file: prl.tex
\documentclass[twocolumn,showpacs,preprintnumbers,aps,prl,superscriptaddress]{revtex4}

\usepackage{amsmath}
\usepackage{amssymb}
\usepackage{graphicx}% Include figure files
\usepackage{dcolumn}% Align table columns on decimal point
\usepackage{bm}% bold math

\newcommand{\BABARPubYear}    {09}
\newcommand{\BABARPubNumber}  {027}

\newcommand{\SLACPubNumber} {13754}

\newcommand{\beq}{\begin{equation}}
\newcommand{\eeq}{\end{equation}}
\newcommand{\beqn}{\begin{eqnarray}}
\newcommand{\eeqn}{\end{eqnarray}}
\newcommand{\beqns}{\begin{eqnarray*}}
\newcommand{\eeqns}{\end{eqnarray*}}

\newcommand{\intl}{\int\limits}

\newcommand{\ppbar}{p\overline{p}}
\newcommand{\mumu}{\mu^+\mu^-}
\newcommand{\pipi}{\pi^+\pi^-}

\input babarsym

\begin{document}

\begin{flushleft}
\babar-PUB-\BABARPubYear/\BABARPubNumber\\
SLAC-PUB-\SLACPubNumber\\
%arXiv:\LANLNumber\ [hep-ex]\\[10mm]~ \\
\end{flushleft}

\title{Precise Measurement of the $e^+e^-\to\pi^+\pi^-(\gamma)$ 
Cross Section with the Initial State Radiation Method at \babar}

\input authors_jul2009_bad2219

\date{\today}

\begin{abstract}
A precise measurement of the cross section of the process 
$e^+e^-\to\pi^+\pi^-(\gamma)$ from threshold to an energy of $3\gev$ 
is obtained with the initial state radiation (ISR) method
using $232\invfb$ of data collected with the \babar\ detector at
$e^+e^-$ center-of-mass energies near $10.6\gev$. 
The ISR luminosity is determined from a study of the 
leptonic process $e^+e^-\to\mu^+\mu^-\gamma(\gamma)$.
The leading-order hadronic contribution to the muon magnetic anomaly
calculated using the $\pi\pi$ cross section measured
from threshold to $1.8\gev$ is
$(514.1 \pm 2.2({\rm stat}) \pm 3.1({\rm syst}))\times 10^{-10}$.
\end{abstract}


\maketitle

Measurements of the $e^+e^-\to{\rm hadrons}$ cross section 
are necessary to evaluate dispersion integrals for calculations of 
hadronic vacuum polarization (VP). Of particular interest is the
contribution $a_\mu^{had}$ to the muon magnetic moment anomaly
$a_\mu$, which requires data in a region 
dominated by the process $e^+e^-\to\pi^+\pi^-(\gamma)$.
Comparison of the theoretical and measured~\cite{bnl} values of $a_\mu$
shows a discrepancy of about $3\sigma$ when current
$e^+e^-$ data~\cite{cmd-2,snd,kloe} are used, possibly hinting at new physics.
An approach using $\tau$ decay data corrected for isospin-breaking,
leads to a smaller difference~\cite{newtau}. 

The results on $\pi\pi$ production reported in this Letter 
are obtained with the ISR method~\cite{isr} 
using $e^+e^-$ annihilation events collected  
at a center-of-mass (CM) energy $\sqrt{s}$ near 10.58 GeV.
The cross section for $e^+e^-\to X$ at the reduced energy $\sqrt{s'}=m_X$, 
where $X$ can be any final state, 
is deduced from a measurement of the radiative process 
$e^+e^-\to X\gamma$ where the photon is emitted by the $e^+$ or $e^-$; 
$s'=s(1-2E_\gamma^*/\sqrt{s})$, where $E_\gamma^*$ is the CM energy of the 
ISR photon.
In this analysis, $\sqrt{s'}$ ranges from threshold 
to 3 GeV.
Two-body ISR processes with $X=\pi^+\pi^-(\gamma)$ and
$X=\mu^+\mu^-(\gamma)$ are measured, where the ISR photon is detected
at large angle and the charged particle pair can be
accompanied by a final state radiation (FSR) photon. Obtaining the $\pi\pi$
cross section from the ratio of pion to muon yield reduces significantly 
the systematic uncertainty. The measured muon cross section is compared to the 
QED prediction, and this cross check of the analysis is termed the QED test. 

The $\sqrt{s'}$ spectrum of $e^+e^-\to X\gamma$ events 
is related to the cross section for the process $e^+e^-\to X$ through
\begin{equation}
\label{def-lumi}
  \frac {\mathrm{d}N_{X\gamma}}{\mathrm{d}\sqrt{s'}}~=~\frac {\mathrm{d}L_{ISR}^{eff}}{\mathrm{d}\sqrt{s'}}~
    \varepsilon_{X\gamma}(\sqrt{s'})~\sigma_{X}^0(\sqrt{s'})~,
\end{equation}
where $\varepsilon_{X\gamma}$ is the detection efficiency (acceptance)
determined by simulation with corrections obtained from data, 
and $\sigma_X^0$ is the bare cross section (excluding VP). The measurement 
of $\sigma_{\pi\pi(\gamma)}^0$ uses the effective ISR luminosity 
$dL_{ISR}^{eff}/\mathrm{d}\sqrt{s'}$ provided by the measured mass spectrum of 
$\mu\mu\gamma(\gamma)$ events following Eq.(\ref{def-lumi}) in which 
$\sigma_X^0(\sqrt{s'})$ is the $\mu\mu(\gamma)$ bare cross section computed 
with QED~\cite{fsr}. 
For the QED test, the measurement of $\sigma_{\mu\mu(\gamma)}^0$
uses the effective ISR luminosity definition 
as a product of the $e^+e^-$ integrated luminosity ($L_{ee}$), 
the radiator function~\cite{isr}, the ratio of detection efficiencies 
for the ISR photon in data and simulation 
(not included in $\varepsilon_{X\gamma}$), 
and the VP correction $(\alpha(s')/\alpha(0))^2$. 
The radiator function, determined by the simulation, is the probability to radiate one or several ISR photons
so that the produced final state $X$ (excluding ISR photons) has mass 
$\sqrt{s'}$. 

This analysis is based on $232\invfb$ of data recorded with the \babar\ 
detector~\cite{detector} at the PEP-II asymmetric-energy $e^+e^-$ storage 
rings. Charged-particle tracks are measured with a five-layer double-sided 
silicon vertex tracker (SVT) together with a 40-layer drift chamber (DCH) 
inside a 1.5 T superconducting solenoid magnet. The energy and direction
of photons are measured in the CsI(Tl) electromagnetic calorimeter (EMC).
Charged-particle identification (PID) uses ionization loss $\dedx$
in the SVT and DCH, the Cherenkov radiation detected in a ring-imaging
device (DIRC), and the shower deposit in the EMC ($E_{cal}$) and in the 
instrumented flux return (IFR) of the magnet.

Signal and background ISR processes are simulated with Monte Carlo (MC) event
generators based on Ref.~\cite{eva}. Additional ISR photons are generated with
the structure function method~\cite{struct-fct}, and additional FSR photons 
with {\small PHOTOS}~\cite{photos}. Background events from $e^+e^-\to\qqbar$ 
($q=u,d,s,c$) are generated with {\small JETSET}~\cite{jetset}. The response 
of the \babar\ detector is simulated with {\small GEANT4}~\cite{geant}. 

Two-body ISR events are selected 
by requiring a photon with $E_\gamma^*>3\gev$ and 
laboratory polar angle in the range $0.35-2.4\rad$, 
and exactly two tracks of opposite charge, each with 
momentum $p>1\gevc$ and within the angular range $0.40-2.45\rad$.
If several photons are detected, the ISR photon is chosen to be that with the 
highest $E_\gamma^*$. The charged-particle tracks, required to have at least 
15 hits in the DCH, must originate within
$5\mm$ of the collision axis and extrapolate to DIRC and
IFR active areas which exclude low-efficiency regions.
An additional criterion based on a combination of 
$E_{cal}$ and $\dedx$ reduces electron contamination.

Acceptance and mass-dependent efficiencies for trigger, reconstruction, 
PID, and event selection are computed using the simulation.
The ratios of data and MC efficiencies have been determined from specific 
studies, as described below, and are applied as mass-dependent 
corrections to the MC efficiency. They amount to at most a few 
percent and are known to a few permil level or better.

Tracking and PID efficiencies are determined 
taking advantage of pair production. For tracking studies, two-prong ISR 
candidates are selected on the basis of the ISR photon and one track.  
A kinematic fit yields the expected parameters of the second track. 
The unbiased sample of candidate second tracks  
is used to measure track reconstruction efficiency.
The maximum correlated two-track loss induced by track overlap in the DCH is
0.6\% for pions and 0.3\% for muons. 

Tracks are assigned uniquely to a complete set of PID classes 
using a combination of cut-based and likelihood selectors.
The `$\mu$' class is addressed first by making use of track IFR penetration 
and hit spread distribution, and of the $E_{cal}$ value. 
Tracks failing the `$\mu$' identification are labeled as `$e$' if they satisfy 
$E_{cal}/p>0.8$. The `$K$' class is determined using DIRC information and 
$\dedx$. Remaining tracks are labeled as `$\pi$'. A tighter selection called 
`$\pi_h$' is applied in mass regions where background dominates  
or to create a pure pion test sample.

Efficiencies for PID are measured from pure samples of muon, pion, and kaon 
pairs obtained from $x\overline{x}\gamma$ events where one track is 
selected as `$\mu$', `$\pi_h$', or `$K$' and the other 
is used to probe the PID algorithm. 
The efficiencies are stored, according to momentum and 
position in the IFR or the DIRC. Typical efficiency for `$\mu$'
is 90\%, with 10\% mis-ID as `$\pi$'.
The `$\pi$' efficiency is strongly momentum-dependent because of
mis-ID as `$K$' (1\% at $1\gevc$, reaching 20\% at $6\gevc$),
as `$\mu$' (5-6\%), or as `$e$' (2\%). Correlations between the 
PID efficiencies, due to track overlap, have been observed and parametrized. 
They are largest for muons where the correlated PID loss reaches 
1.3\% of the events below $1\gevcc$. It is important to control this effect,
since it affects the $\pi\pi$ and $\mu\mu$ samples in an anti-correlated way. 

To obtain the spectra $N_{jj}$ of produced particle pairs of true type $j$,
a set of three linear relations must be solved. They involve the
$N_{jj}$, the measured mass distributions for each 
`$ii$'-identified final state, and the 
probabilities $\varepsilon^{jj}_{`ii\textit{'}}$  ($i,j=\mu, \pi$ or $K$) which 
represent the product of the measured efficiencies for each track of true type 
$j$ to be identified as `$i$', corrected by correlation factors.

A contribution ($<10^{-3}$) to $N_{\pi\pi}$ from $\ppbar\gamma$  
is estimated from MC and subtracted after reweighting the rate to 
agree with the \babar\ measurement~\cite{ppb}. Multi-hadronic background 
from $e^+e^-\to\qqbar$ comes from low-multiplicity events 
in which an energetic $\gamma$ 
originating from a $\pi^0$ is mistaken as the ISR photon candidate. 
To normalize this rate from {\small JETSET}, the $\pi^0$ yield obtained by
pairing the ISR photon with other photons in the event  
is compared in data and MC; {\small JETSET} overestimates this background by 
a factor 1.3. Multi-hadronic ISR backgrounds are dominated by
$e^+e^-\to\pi^+\pi^-\pi^0\gamma$ and $e^+e^-\to\pi^+\pi^-2\pi^0\gamma$ 
contributions. An approach similar to that for $\qqbar$ is followed to 
calibrate the background level from the $3\pi$ ISR process,
using $\omega$ and $\phi$ signals.
The ratio of data to MC yield is found to be
$0.99\pm0.04$. The MC estimate for the $2\pi 2\pi^0\gamma$ process is
used and assigned a 10\% systematic uncertainty.
A residual radiative Bhabha background is identifiable only near 
threshold and at large mass, where the pion signal vanishes. Its magnitude 
is estimated from the helicity angle distribution in the $\pi\pi$ CM frame
at low energy and its energy dependence obtained from a control sample of 
radiative Bhabha events. It is assigned a 100\% systematic uncertainty. 
To suppress the contribution from the 
$e^+e^-\to\gamma\gamma$ process with a photon conversion,
which affects the spectrum at threshold, 
the vertex of the two tracks is required to be closer than $5\mm$ to the 
collision axis in the transverse plane. This criterion is applied only 
to events in the $\rho$ tails, defined to lie outside the 
central region $0.5<m_{\pi\pi}<1.0\gevcc$.
Background contributions to the $N_{\mu\mu}$ spectrum are negligible.

Each event is subjected to two kinematic fits to the $e^+e^-\to X\gamma$ 
hypothesis,
where $X$ includes one additional photon, detected or not.
Both fits use the ISR photon direction and the parameters and
covariance matrix of each charged-particle track.
The energy of the ISR photon is not used, as it 
has little impact for the relatively low CM energies involved.
The two-constraint (2C) `ISR' fit allows an undetected photon 
collinear with the collision axis, while the 3C `FSR' fit uses any 
photon with $E_{\gamma}>25\mev$. When more than one such photon is present, 
the best `FSR' fit is retained. An event with no extra photon is characterized 
only by its $\chi^2_{ISR}$ value.
Most events have small $\chi^2$ values for both fits; an event with only a
small $\chi^2_{ISR}$ ($\chi^2_{FSR}$) indicates the presence of additional 
ISR (FSR) radiation. Events where both fits have large $\chi^2$ values 
result from track or ISR photon resolution effects, the presence of 
additional radiated photons, or multi-hadronic background.
To accommodate the expected background levels, different criteria in the 
($\chi^2_{ISR}$,$\chi^2_{FSR}$) plane are applied depending on the 
$m_{\pi\pi}$ mass regions. For the central $\rho$ region,
a loose 2D contour has been optimized to remove the main background area 
while maintaining control of the associated systematic uncertainties.
The same procedure is used in the $\mu\mu\gamma$ analysis in spite of 
the very small background. In the $\rho$ tails, a tighter $\chi^2$ selection 
is imposed to reduce the larger background. 
Samples of 529320 pion and 445631 muon events are selected in the mass range 
below $3\gevcc$, where the $m_{\pi\pi}$ ($m_{\mu\mu}$) mass is calculated 
from the best `ISR' or `FSR' fit.

The computed acceptance and the $\chi^2$ selection efficiency 
depend on the description of radiative effects in 
the generator. The FSR rate is measured from events that satisfy the `FSR' fit,
with an additional photon ($E_{\gamma}>0.2\gev$) within $20\degrees$ of either
track. The excess in data relative to the generator prediction 
using {\small PHOTOS}~\cite{photos} is $(-4\pm6)$\% of total FSR for muons, 
and $(21\pm5)$\% for pions. This difference results in a 
$(6\pm2)\times 10^{-4}$ correction. More significant differences are found
between data and the generator for additional ISR photons, since the latter
uses a collinear approximation and an energy cut-off for very hard photons. 
Induced kinematical effects have been studied using the next-to-leading order 
(NLO) {\small PHOKHARA} generator~\cite{phok} at four-vector
level with fast simulation. Differences in acceptance occur at the few 
percent level, and these yield corrections to the QED test. In contrast, 
since radiation from the initial state is common to the pion and muon
channels, the $\pi\pi(\gamma)$ cross section, obtained from the
$\pi\pi$/$\mu\mu$ ratio, is affected and corrected only at a few permil level.
Additional ISR effects on the $\chi^2$ selection efficiencies factorize 
in both processes and cancel in the ratio.
The $\chi^2$ selection efficiency determined from muon data 
applies to pions, after correction for the effect of secondary
interactions and the $\pi/\mu$ difference for additional FSR. 
Therefore the measurement of the pion 
cross section is to a large extent insensitive to the description of NLO 
effects in the generator.

\begin{figure}[htp]
  \centering
  \includegraphics[width=9cm]{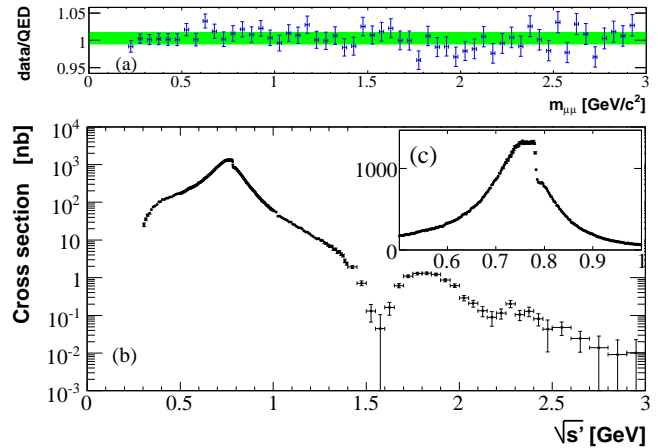}
  \caption{\small
(a) The ratio of the measured cross section for 
$e^+e^-\to\mu^+\mu^-\gamma(\gamma)$ to the NLO QED prediction. 
The band represents Eq.~(\ref{qed-test}).
(b) The measured cross section for $e^+e^-\to\pi^+\pi^-(\gamma)$ 
from 0.3 to $3\gev$. 
(c) Enlarged view of the $\rho$ region in energy intervals of 2 MeV.
The errors are from the combined diagonal elements of the statistical and 
systematic covariance matrices.}
  \label{babar-log}
\end{figure}

The QED test involves two additional factors, both of which 
cancel in the $\pi\pi$/$\mu\mu$ ratio: $L_{ee}$ and the ISR photon efficiency, 
which is measured using a $\mu\mu\gamma$ sample selected only on the basis 
of the two muon tracks. The QED test is expressed as the ratio  
of data to the simulated spectrum, after the latter is corrected using data 
for all known detector and reconstruction differences. The generator is 
also corrected for its known NLO deficiencies using the comparison to 
{\small PHOKHARA}. The ratio is consistent with unity from threshold to 
$3\gevcc$, (Fig.~\ref{babar-log} (a)). A fit to a constant value yields 
($\chi^2/n_{\rm{df}}=55.4/54$; $n_{\rm{df}}$=number of degrees of freedom)
\begin{equation}
\label{qed-test}
 \frac {\sigma_{\mu\mu\gamma(\gamma)}^{data}} {\sigma_{\mu\mu\gamma(\gamma)}^{NLO~QED}}~-~1~=~ (40\pm20\pm55\pm94)\times 10^{-4}~,
\end{equation}
where the errors are statistical, systematic from
this analysis, and systematic from $L_{ee}$, respectively.
The QED test is thus satisfied within an overall accuracy of 1.1\%.

To correct for resolution and FSR effects, an unfolding of the 
background-subtracted and efficiency-corrected
$m_{\pi\pi}$ distribution is performed.
A separate mass-transfer matrix is created using simulation for the 
$\rho$ central and tail regions;
this provides the probability that an event generated in a $\sqrt{s'}$ 
interval $i$ is reconstructed in a $m_{\pi\pi}$ interval $j$. 
The matrix is corrected using 
data to account for the larger rate of events with poorer mass resolution.  
Performance and robustness of the unfolding method~\cite{bogdan} 
have been assessed using test models. For the 2-MeV intervals,
the significant elements of the resulting covariance matrix lie near the 
diagonal over a typical range of $6-8\mev$, which corresponds to the 
energy resolution. 

\begin{table} [b] \centering 
\caption{ \label{syst-err} \small 
Relative systematic uncertainties (in $10^{-3}$) 
on the $e^+e^-\to\pi^+\pi^-(\gamma)$ 
cross section by $\sqrt{s'}$ intervals (in $\gev$) up to $1.2\gev$. 
The statistical part of the efficiency uncertainties is included in 
the total statistical uncertainty in each interval.}
\vspace{0.5cm}
\begin{tabular}{|c|ccccc|} \hline\hline
Source of      & \multicolumn{5}{c|}{CM Energy Interval (GeV)}\\
\cline{2-6}
 Uncertainty      &  0.3-0.4 & 0.4-0.5 & 0.5-0.6 & 0.6-0.9 & 0.9-1.2 \\ \hline
 trigger/ filter            & 5.3 & 2.7 & 1.9 & 1.0 & 0.5  \\ 
 tracking                   & 3.8 & 2.1 & 2.1 & 1.1 & 1.7  \\ 
 $\pi$-ID                   &10.1 & 2.5 & 6.2 & 2.4 & 4.2  \\
 background                 & 3.5 & 4.3 & 5.2 & 1.0 & 3.0  \\
 acceptance                 & 1.6 & 1.6 & 1.0 & 1.0 & 1.6  \\
 kinematic fit ($\chi^2$)   & 0.9 & 0.9 & 0.3 & 0.3 & 0.9 \\
 correlated $\mu\mu$ ID loss     & 3.0 & 2.0 & 3.0 & 1.3 & 2.0 \\
 $\pi\pi/\mu\mu$ non-cancel.& 2.7 & 1.4 & 1.6 & 1.1 & 1.3 \\
 unfolding                  & 1.0 & 2.7 & 2.7 & 1.0 & 1.3  \\
 ISR luminosity ($\mu\mu$)  & 3.4 & 3.4 & 3.4 & 3.4 & 3.4 \\
\hline
 total uncertainty          &13.8 & 8.1 &10.2 & 5.0 & 6.5  \\
\hline\hline
\end{tabular}
\end{table}

The results for the $e^+e^-\to\pi^+\pi^-(\gamma)$ bare cross 
section~\cite{epaps} including FSR, $\sigma^0_{\pi\pi(\gamma)}(\sqrt{s'})$, 
are given in Fig.~\ref{babar-log} (b).
Prominent features are the dominant $\rho$ resonance, the abrupt drop at 
$0.78\gev$ due to $\rho-\omega$ interference, a clear dip at $1.6\gev$ 
resulting from higher $\rho$ state interference,
and additional structure near $2.2\gev$.
Systematic uncertainties are estimated from the precision of the data-MC 
comparisons and from the measurement procedures used for the various 
efficiencies. They are reported in Table~\ref{syst-err} for 
$0.3<\sqrt{s'}<1.2\gev$. Although larger outside this range, the systematic 
uncertainties do not exceed statistical errors over the full spectrum for the
chosen energy intervals.

The lowest-order contribution of the $\pi\pi(\gamma)$ intermediate state
to the muon magnetic anomaly is given by 
\begin{equation}
\label{eq:int_amu}
    a_\mu^{\pi\pi(\gamma),LO} \:=\: 
       \frac{1}{4\pi^3}\!\!
       \intl_{4m_\pi^2}^\infty\!\!\mathrm{d}s'\,K(s')\,\sigma^{0}_{\pi\pi(\gamma)}(s')~,
\end{equation}
where $K(s')$ is a known kernel~\cite{kernel}.
The integration uses the measured cross section and the
errors are computed using the full statistical and systematic covariance 
matrices. The systematic uncertainties for each source are taken to be fully 
correlated over all mass regions. 
The integrated result from threshold to $1.8\gev$ is
\begin{equation}
    a_\mu^{\pi\pi(\gamma),LO} \:=\: (514.1 \pm 2.2 \pm 3.1)\times 10^{-10}~,
\end{equation}
where the errors are statistical and systematic.
This value is larger than that from a combination of previous 
$e^+e^-$ data~\cite{newtau} ($503.5\pm3.5$), but is 
in good agreement with the updated value from $\tau$ decay~\cite{newtau} 
($515.2\pm3.4$).

In summary, the cross section for the process $e^+e^-\to\pi^+\pi^-(\gamma)$ 
has been measured in the energy range from 0.3 to $3\gev$, 
using the ISR method. The result for the $\pi\pi$ hadronic contribution to 
$a_{\mu}$ has a precision comparable to that of the combined
value from existing $e^+e^-$ experiments. However, the \babar\ central value 
is larger, which reduces the deviation of the direct $a_\mu$
measurement from the Standard Model prediction.

\input acknow_PRL

\end{document}

%% file: authors_jul2009_bad2219.tex
%% author list as of 10-Jul-2009 (484 authors)
%
\author{B.~Aubert}
\author{Y.~Karyotakis}
\author{J.~P.~Lees}
\author{V.~Poireau}
\author{E.~Prencipe}
\author{X.~Prudent}
\author{V.~Tisserand}
\affiliation{Laboratoire d'Annecy-le-Vieux de Physique des Particules (LAPP), Universit\'e de Savoie, CNRS/IN2P3,  F-74941 Annecy-Le-Vieux, France}
\author{J.~Garra~Tico}
\author{E.~Grauges}
\affiliation{Universitat de Barcelona, Facultat de Fisica, Departament ECM, E-08028 Barcelona, Spain }
\author{M.~Martinelli$^{ab}$}
\author{A.~Palano$^{ab}$ }
\author{M.~Pappagallo$^{ab}$ }
\affiliation{INFN Sezione di Bari$^{a}$; Dipartimento di Fisica, Universit\`a di Bari$^{b}$, I-70126 Bari, Italy }
% ...
\collaboration{The \babar\ Collaboration}
\noaffiliation

%% file: acknow_PRL.tex
We are grateful for the excellent luminosity and machine conditions
provided by our \pep2\ colleagues, 
and for the substantial dedicated effort from
the computing organizations that support \babar.
The collaborating institutions wish to thank 
SLAC for its support and kind hospitality. 
This work is supported by
DOE
and NSF (USA),
NSERC (Canada),
CEA and
CNRS-IN2P3
(France),
BMBF and DFG
(Germany),
INFN (Italy),
FOM (The Netherlands),
NFR (Norway),
MES (Russia),
MEC (Spain), and
STFC (United Kingdom). 
Individuals have received support from the
Marie Curie EIF (European Union) and
the A.~P.~Sloan Foundation.

%% file: prl.bbl
\begin{thebibliography}{99}
\bibitem{fsr} Correction for the effect of lowest-order FSR is applied.
\end{thebibliography}
